\documentclass[a4paper,11pt]{article}
\pdfoutput=1 

\usepackage{jinstpub} 

\usepackage{hyperref}
\title{Development of gamma insensitive silicon carbide diagnostics to qualify intense thermal and epithermal neutron fields}

\author[a,b,1]{O. Sans Planell,\note{Corresponding author.}}
\author[a,b]{M. Costa,}
\author[a,b]{E. Durisi,}
\author[a,c]{A. Lega,}
\author[a,b]{E. Mafucci,}
\author[a,c]{L. Menzio,}
\author[a,b]{V. Monti,}
\author[a,b]{L. Visca,} 
\author[c]{R. Bedogni,}
\author[d]{M. Treccani,}
\author[e,f]{A. Pola,}
\author[e,f]{D. Bortot,}
\author[g,h]{K. Alikaniotis,}
\author[g,h]{G. Giannini,}
\author[i]{J. M. Gomez-Ros.}

\affiliation[a]{Universita degli Studi di Torino,\\Via Pietro Giuria 1, 10125, Torino, Italy}
\affiliation[b]{INFN, Sezione di Torino,\\Via Pietro Giuria 1, 10125, Torino, Italy}
\affiliation[c]{INFN, Laboratori Nazionali di Frascati,\\Via Enrico Fermi 40, 00044, Frascati, Italy}
\affiliation[d]{Universitat Autonoma de Barcelona, Departament de Fisica,\\08193, Bellaterra, Spain}
\affiliation[e]{Politecnico de Milano, Dipartimento di Energia,\\Via La Masa 34, 20156, Milano, Italy}
\affiliation[f]{INFN, Sezione di Milano,\\Via La Masa 34, 20156, Milano, Italy}
\affiliation[g]{Universita degli Studi di Trieste,\\Via Valerio 2, 34127, Trieste, Italy}
\affiliation[h]{INFN, Sezione di Trieste,\\Via Valerio 2, 34127, Trieste, Italy}
\affiliation[i]{CIEMAT,\\Avenida Complutense 40, 28040, Madrid, Spain}

\emailAdd{oriol.sansplanell@to.infn.it}

\abstract{The e\_LiBANS project aims at creating accelerator based compact neutron facilities for diverse interdisciplinary applications.
After the successful setting up and characterization of a thermal neutron source based on a medical electron LINAC, a similar assembly for epithermal neutrons has been developed. The project is based on
an Elekta 18 MV LINAC coupled with a photoconverter-moderator system
which deploys the ($\gamma$,n) photonuclear reaction to convert a bremsstrahlung
photon beam into a neutron field. This communication describes the development of novel diagnostics to qualify the thermal and epithermal neutron fields that have been produced. In particular, a proof of concept for the use of silicon carbide photodiodes as thermal neutron rate detector is presented.}


\proceeding{15$^{\text{th}}$ Topical Seminar on Innovative Particle and Radiation Detectors\\
  14-17 October 2019\\
  Siena (Italy)}

\begin{document}
\maketitle
\flushbottom

\section{Introduction}
\label{sec:intro}
A worldwide interest exists to develop compact and cost-wise low energy neutron sources. The most common ones are based on proton or deuteron beams on different materials \cite{Carpenter} with relevant technological and cost-effective investments. An alternative to such sources is the use of the $(\gamma,n)$ reaction, in which neutrons with a typical evaporation spectrum around 1-2 MeV are produced, by converting the high energy photons on a heavy material target. Exploiting a modified medical LINAC, the INFN e\_LiBANS collaboration has successfully developed two converter-moderator assemblies, one for thermal neutrons and a second one for epithermal neutrons.  
\newline
The e\_LIBANS project started back in 2016 and ended in spring 2019. For an extensive description see \cite{ValeThesis}. It offers two fully functional facilities with a wide range of applications: from cell testing for medical research, to detector characterisation. Alongside the creation of the two photoconverters-moderators, some novel diagnostics have been built to properly measure the neutron fluence rates and the energy spectra. The root technology, upon which the new diagnostics are based, is the Thermal Neutron Rate Detector (TNRD) developed by the group under previous projects \cite{TNRD}.
The "TNRD technology" is characterized by the differential readout of an assembly of two silicon detectors where a thin layer of  ${}^6$LiF is deposited on one of the two detectors while the other operates "bare". The ${}^6$LiF has an extremely high absorption cross section for thermal neutrons. When the ${}^6$Li nucleus absorbs a neutron, it undergoes the following nuclear reaction: 
\begin{equation}
    n + {}^6Li \xrightarrow{} \alpha + {}^3H + 4.75 MeV
    \label{eq:lithium}
\end{equation}
The geometry of the detector assembly is optimized  so that the intrinsically short-range products of the nuclear reaction leave their signal only in the deposited diode, while background photons can produce signals in both devices.  A differential readout has been implemented, so that the photon signal tend to cancel out, leaving the neutron signal intact. This method makes the device virtually insensitive to photons and, thus, optimal to work in a mixed $\gamma$-n field. The detector has shown a linear response on a wide neutron fluence-rate range and it has also been proved to work well under pulsed fields, such as that of a LINAC. The radiation hardness of the silicon substrate of the TNRD has proved to withstand neutron fluences up to $5x10^{11}$ cm$^{-2}$ \cite{TNRD1}, deeming it adequate for radioprotection purposes, although it has inspired a research for more radiation resistant devices. In this paper two new detectors are presented: the first deals with the implementation of the TNRD technology on a silicon carbide substrate to reach higher radiation hardness, while the second one aims at extending the "TNRD technology" to a fluence-meter device able to operate in intense epithermal neutron fields. 
\newline
For the determination of the neutron energy spectra, a Bonner Sphere Spectrometer has been used, coupled to a TNRD as active detector. The TNRD-BSS spectrometer has been calibrated at the National Physical Laboratory (England) and at the ENEA-Frascati Neutron Generator (Italy \cite{Tonino}).

\section{The e\_LiBANS testing facilities}
The e\_LiBANS collaboration can rely on three different calibrated neutron facilities that have been setup by the group. Some details are given in the following sections. 

\subsection{The e-Linac based facility in Torino}
\begin{figure}[hb]
    \centering
    \includegraphics[width=.9\textwidth]{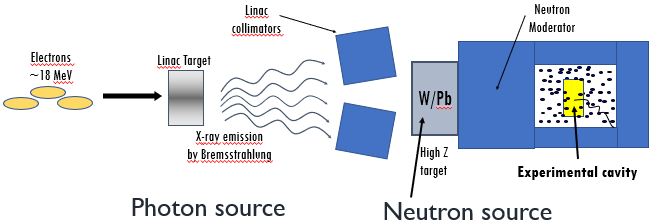}
    \caption{Schematic view of the different components of the e\_LiBANS neutron source.}
    \label{fig:photoproduction}
\end{figure}

\begin{figure}[htbp]
\centering 
\includegraphics[width=.9\textwidth]{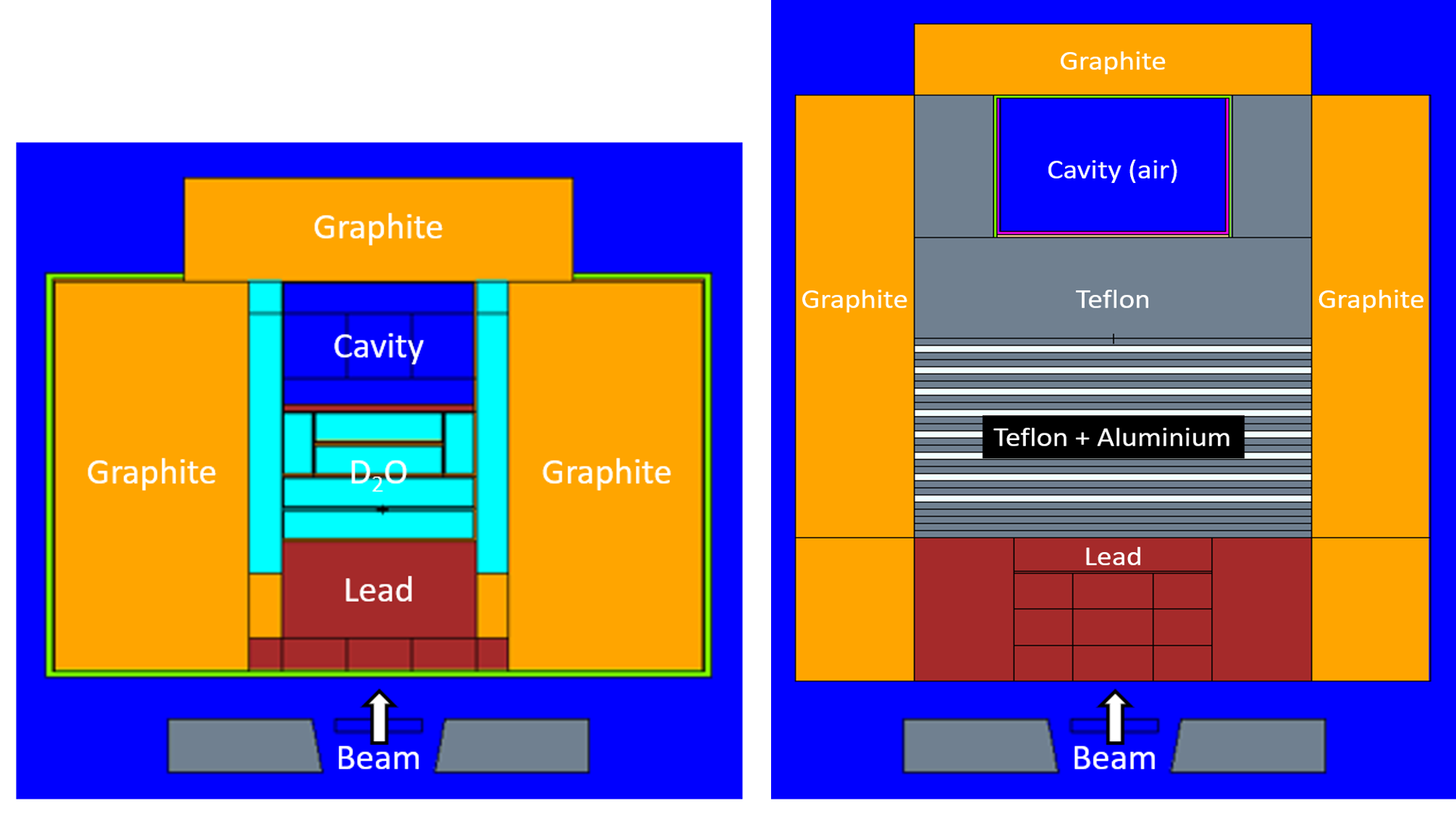}
\caption{\label{fig:i} {\bf Left}: the thermal neutron photoconverter-moderator geometry. {\bf Right}: the epithermal neutron photoconverter-moderator geometry. The beam direction is also indicated in both figures.}
\end{figure}

The Torino facility exploits the bremsstrahlung gammas produced by the conversion of the electron beam, produced by an ELEKTA SL 18 MeV Linac, on a thin tungsten target \cite{elibans_vale}.
The high energy gammas can then be used to produce neutrons through $(\gamma,n)$ reaction on a high-Z material as lead. Neutrons are produced with a typical evaporation spectrum peaked around 1 -2 MeV. The thickness of the target can be tuned to maximise both the conversion process, and the absorption of the unconverted gammas. The processes are represented in figure\ref{fig:photoproduction}.
\newline
After being extracted from the target, neutrons pass through a region in which they undergo moderation (i.e. reduction of their energy) to the desired range. Two different moderator assemblies were constructed in order to obtain homogeneous neutron fields with energy in the thermal and epithermal ranges, respectively.
\newline
In the thermal configuration (see figure \ref{fig:i}), the producer block is surrounded by an external graphite structure, followed by a central core of deuterated water that thermalizes the neutrons. The minimal absorption cross section and the high scatter capabilities of those materials makes them ideal for reflection and moderation. A coating made of borated rubber and polyethylene highly reduces the dose imparted to the exterior of the photo-converter. Embedded in the moderator can be found an experimental cavity of 30x30x20 cm$^3$ volume in which it is possible to place samples and detectors. The overall structure has a volume slightly above 1 m$^3$ and it weights 1 Ton \cite{elibans_vale}.
\newline
In the epithermal configuration, the lead target design is similar to that of the thermal structure, but the materials for the moderation vary: in order to keep the neutron energy confined in the epithermal range and to eliminate the thermal and fast components of the spectrum, a combination of aluminium and polytetrafluoroethylene (PTFE) is used as core, while a thin shield of borated rubber - 0.5 cm - and cadmium has been placed surrounding the internal walls of the cavity to eliminate the thermal component of the spectrum.
All the structures have been optimised through extensive Monte Carlo simulation work with MCNP6 \cite{MCNP6} as transport code. The cross section data library used to work out the calculations was the ENDF-B/VII.1 \cite{ENDF}. The configurations are graphically described on figure \ref{fig:i}.
\newline
The Torino neutron facilities underwent extensive characterization and calibration campaigns. Typical neutron fluence rates are of the order of $10^5 -10^6$ cm$^{-2}$s$^{-1}$ in the thermal energy range and $10^4 - 10^5$ cm$^{-2}$s$^{-1}$ in the epithermal energy range. 
As far as the gamma background is concerned the gamma dose rate measured in the centre of the cavity results:
D$_{\gamma} = (1.85\pm0.08)$ ${\mu}$Gy s$^{-1}$. For more details see \cite{elibans_vale}
\subsection{The HOTNES and EPINES facilities}
\label{epi_hot}

\begin{figure}[h]
    \centering
    \includegraphics[scale=0.5]{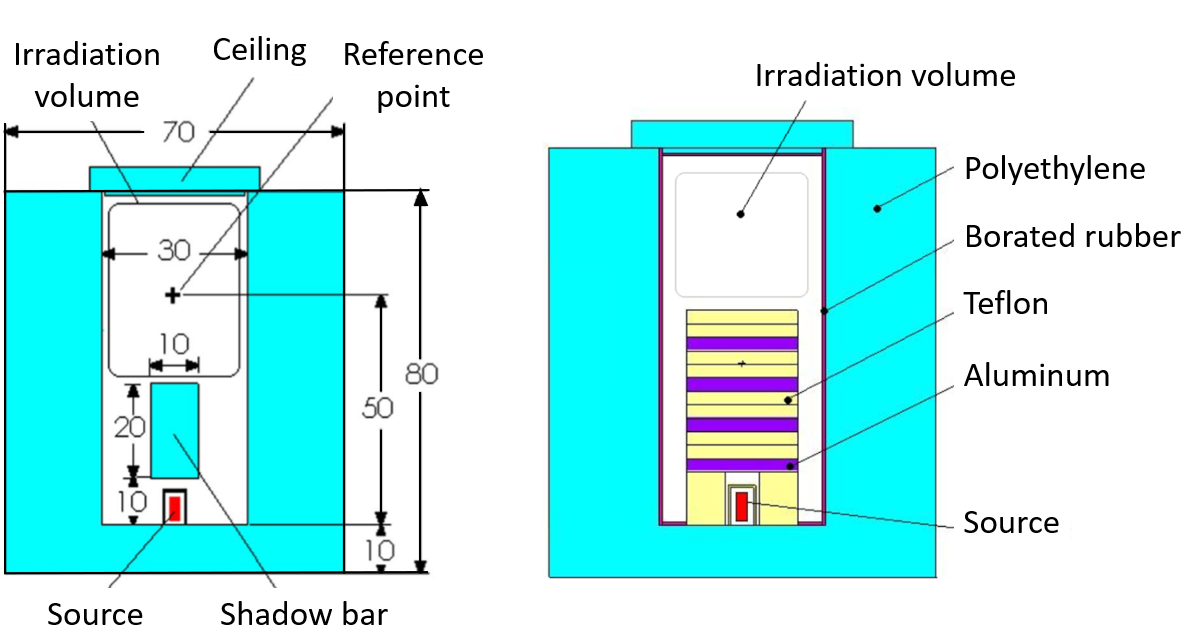}
    \caption{{\bf Left}: Details of HOTNES. {\bf Right}: Schematics of EPINES.}
    \label{fig:hotepi}
\end{figure}

Two more detector testing facilities are available in Frascati to the ANET collaboration, that have been built thanks to a collaboration between ENEA and INFN: the HOmogueneus Thermal NEutron Source (HOTNES \cite{hotnes0}) and the EPIthermal  NEutron  Source (EPINES). Both of them exploit a Am-B neutron source and provide a uniform neutron field in a specific volume, with almost zero gamma contamination, although with limited intensity with respect to the Torino e-Linac sources.  An accurate characterisation of such field was carried out employing a calibrated set of Bonner Spheres \cite{hotnes1}. Thanks to the large accessible volume in which is possible to place the detectors and the uniformity of the neutron field provided, these two sources are ideal for test and calibration of novel detectors. In HOTNES and EPINES it is possible to test, respectively, thermal and epithermal neutron detectors. The simulated geometries for both facilities are presented in figure \ref{fig:hotepi}.
\newline
The work illustrated in this paper has profited of all the facilities previously described.  

\section{Novel diagnostics}

\subsection{Silicon Carbide Neutron Detectors}

\begin{figure}[htbp]
\centering 
\includegraphics[width=.45\textwidth]{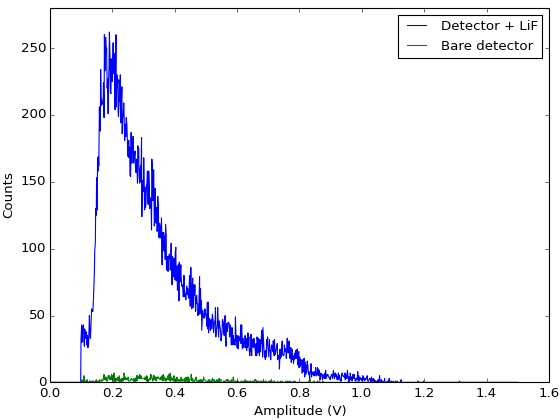}
\qquad
\includegraphics[width=.45\textwidth,origin=c]{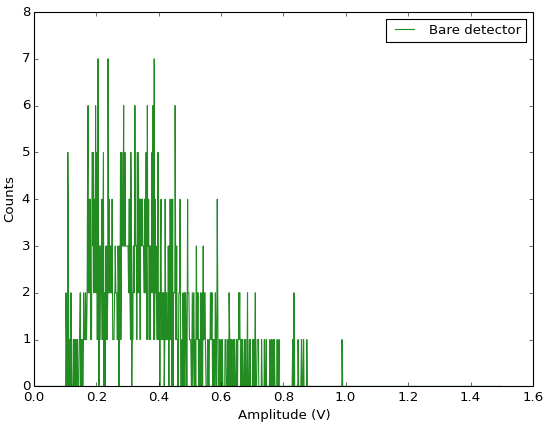}
\caption{\label{fig:SiC_compare} Comparison between the signal pulse heights measured by two SG01XL SiC detectors exposed to the e\_LiBANS thermal neutron field. For clarity, the result with the bare detector is also replicated in the picture on the right.}
\end{figure}

\begin{figure}[htbp]
    \centering 
    \includegraphics[width=.45\textwidth]{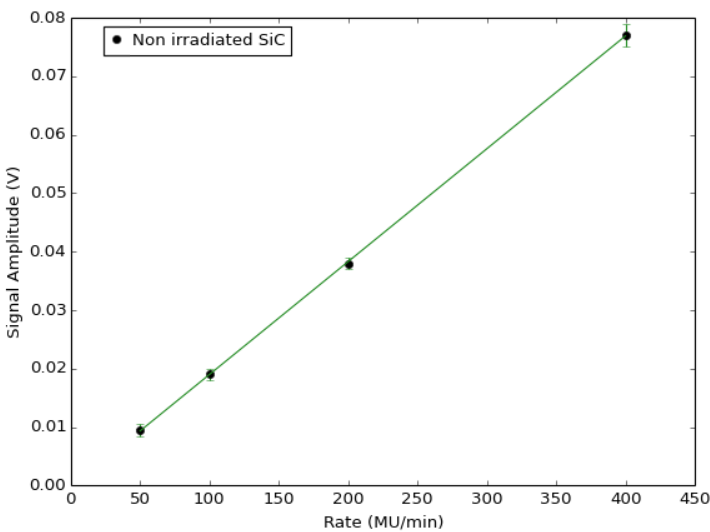}
    \qquad
    \includegraphics[width=.45\textwidth,origin=c]{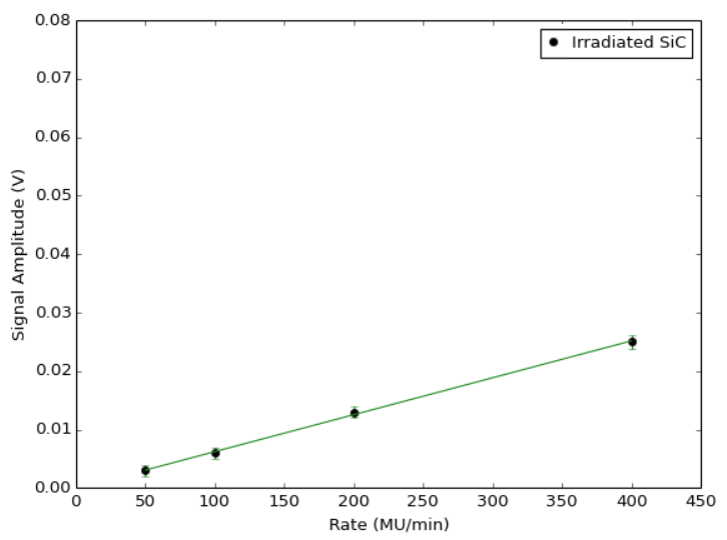}
    \caption{ {\bf Left}: Linearity test performed with the SiC detector under the thermal field in Turin, at rates ranging from 50 to 400 Monitor-Units(MU)/min. {\bf Right}: The same measurement performed on an irradiated detector exposed to an integrated neutron fluence of $10^{13} $cm$^{-2}$.}
    \label{fig:SiC_rates}
\end{figure}

\begin{figure}[h]
    \centering
    \includegraphics[width=0.35\textwidth]{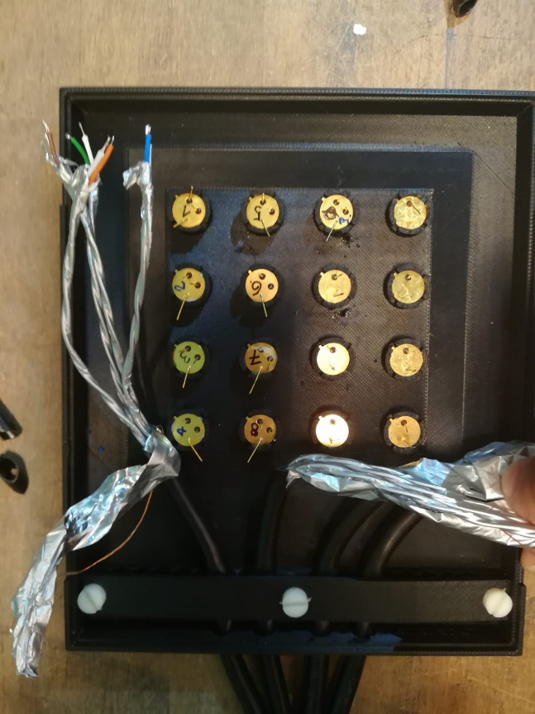}
    \qquad
    \includegraphics[width=0.55\textwidth]{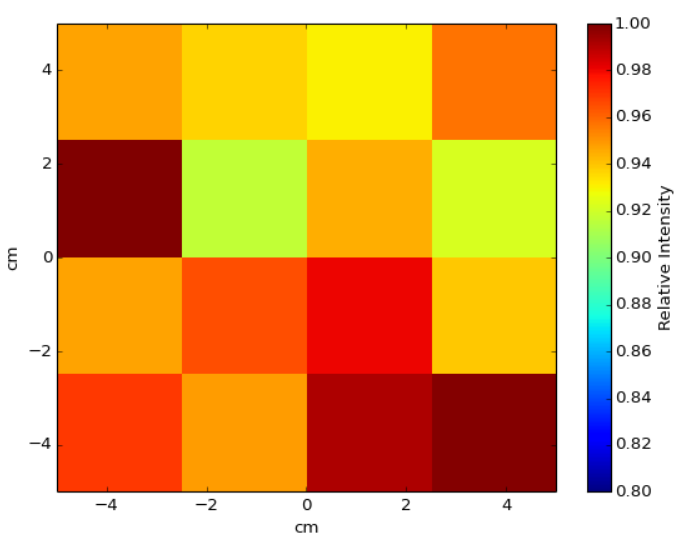}
    \caption{{\bf Left}: Matrix of Silicon Carbide photodiodes for the e$\_$LiBANS project. {\bf Right}: Measurement performed with the matrix inside the e$\_$LiBANS thermal cavity.}
    \label{fig:my_label_1}
\end{figure}

One of the main concerns about the silicon detectors used for the TNRDs resided in their limited radiation hardness. As stated in \cite{TNRD1} their signal started to be compromised for neutron fluences above 5x$10^{11}$ cm$^{-2}$ . In order to find more radiation hard substrates, the behaviour of some commercial silicon carbide (SiC) devices was inspected. This material constitutes a promising choice for high fluence thermal neutron fields, as its energy gap is about three times greater than silicon's. Our choice focused on SGLux GmBh photodiodes \cite{SiC}, originally conceived for UV measurements. They are extremely low noise devices with dark current in the fA range. They can have active areas ranging from 1 mm$^2$ to 7.6 mm$^2$: this allows to scale the geometry of the device to the source fluence rate, in order to get manageable counting rates.
Moreover the junction contact potential provides around 1 $\mu$m depletion layer thickness at zero external bias, that can be exploited to detect alphas and tritii from neutron conversions, making the device nearly insensitive to photons. In unbiased operative conditions the relative gamma to neutron sensitivity has been evaluated to be $10^{-4}$. In the following, results with a 7.6 mm$^2$ SG01XL device by SGLux GmBh are shown. Firstly we compared the response to thermal neutrons of a bare SiC detector with respect to one coated with a thin layer of ${}^6$LiF. Measurements have been taken at the e\_LiBANS thermal facility in Torino. Results are shown in figure \ref{fig:SiC_compare}.
\newline
The bare detector has about $1\%$ counts of the ${}^6$LiF coated one. These few counts are due to the boron doping of the SiC junction and can be assumed to give a negligible contribution to the total signal. This also suggests the possibility to use a linear readout, instead of a differential one as it was applied for the TNRDs.
\newline
The linearity of ${}^6$LiF coated SG01XL SiC detectors has also been studied. Results are shown in figure \ref{fig:SiC_rates}, where the average of the signal amplitude, readout in current-mode, as a function of the Linac beam rate is shown. The maximum rate corresponds to a thermal neutron fluence rate of $2 x 10^6$cm$^{-2}$s$^{-1}$. The device linearity has been proved to be extremely good and its capability to cope with intense neutron rate has been demonstrated.
\newline
To check  the ${}^6$LiF coated SG01XL SiC radiation hardness, we exposed the device to a total neutron fluence of $10^{13} $cm$^{-2}$ at the ENEA Casaccia TRIGA reactor \cite{TRIGA}. This fluence value was chosen as it would represent a three years exposure in the e\_LiBANS thermal neutron facility at maximumm rate.  We then repeated the linearity test at the Linac facility, as previously described. The result is shown in figure \ref{fig:SiC_rates}. 
\newline
As expected the charge collection efficiency got affected by the high radiation dose: a reduction of about a factor three is observed. Nonetheless the linearity of the response remains unchanged and the signal amplitude is still sufficient to operate the device. We observe a net improvement with respect to silicon based devices \cite{TNRD}. The response degradation suggests the need of a periodical calibration to adjust the detector response. More radiation campaigns on SGLux GmBh SiC devices are needed to come to a conclusive statement on their radiation hardness.
\newline
Nonetheless, given their good response, linearity and gamma insensitivity, 16 ${}^6$LiF coated SG01XL SiC detectors were assembled into a 4 x 4 matrix (so called SiC-Matrix), as shown in figure \ref{fig:my_label_1}, to build a beam monitor detector with a field of view of 10 x 10 cm$^2$, with the capacity of being readout simultaneously in both impulse or current modes. 
The uniformity of the field inside the e\_LiBANS thermal cavity was measured with the SiC-Matrix beam monitor. The individual SiC response is known within 5\% error.
The uniformity result is shown in figure \ref{fig:my_label_1} and it is in agreement with what has been previously quoted \cite{elibans_vale}

\subsection{Epithermal Neutron Rate Detector}

\begin{figure}[h]
    \centering
    \includegraphics[scale=0.62]{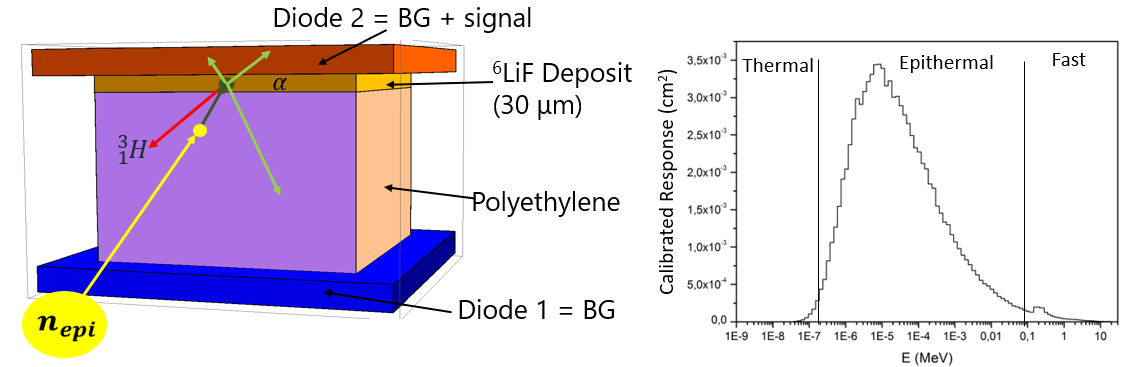}
    \caption{\textbf{Left}: the design of the epithermal neutron rate detector. \textbf{Right}: the response curve of the detector as a function of the energy.}
    \label{fig:EPI3}
\end{figure}

\begin{figure}[htbp]
    \centering 
    \includegraphics[width=.6\textwidth]{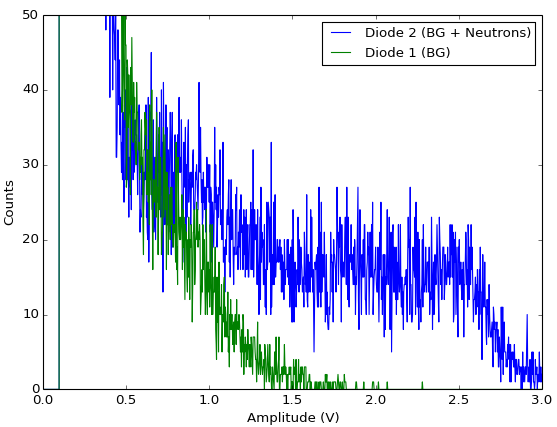}

    \caption{ Measurement taken with the EPI3 detector  at the epithermal facility in Turin. The two diode signals are shown separately.}
    \label{EPI_signal}
\end{figure}

\begin{figure}[htbp]
    \centering
    \includegraphics[width=.5\linewidth]{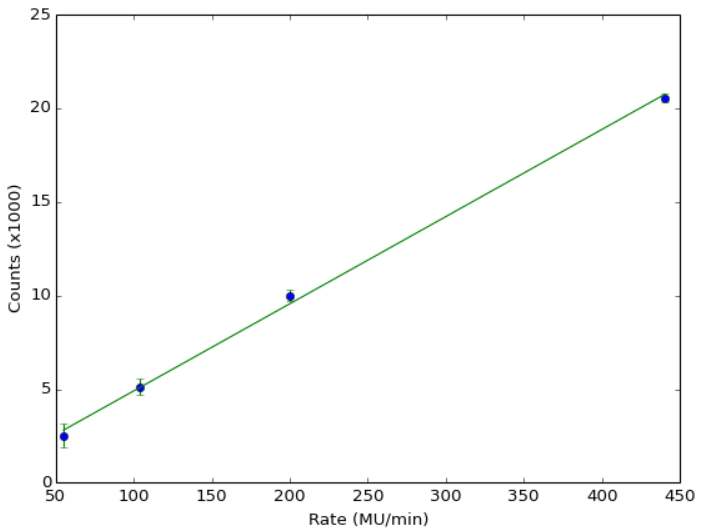}
    \caption{Counts measured by the EPI3 detector as a function of the e-LINAC rate.}
    \label{fig:EPI3_linearita}
\end{figure}

Basing its structure on the presented TNRD, an Epithermal Neutron Rate Detector (labeled EPI3), was build by placing a cube of high density polyethylene between two silicon photodiodes, one of which (Diode2) was coated with a thin layer of ${}^6$LiF, while the other (Diode1) remained bare without any coating, as shown in figure \ref{fig:EPI3}. The whole structure is enclosed in a box of borated rubber that has been dimensioned to absorb neutron in the thermal energy range. Those above the epithermal threshold (0.4 eV) penetrate the borated cup and get thermalized along their path in the polyethilene, before getting converted in the ${}^6$LiF silicon coating of Diode2. The differential readout of the diodes gets rid of the possible gamma contribution and provides a clean neutron signal. The result is an effective, compact and active device to measure epithermal neutron fluence rate. The prototype discussed in this paper has dimensions 4.5 x 3 x 2 cm$^3$.  
The EPI3 response function has been simulated using the MCNP6 code and the result is shown in figure \ref{fig:EPI3} (right).
\newline
The EPI3 detector has been calibrated at the EPINES facility by comparing its response to that extracted from the measurement with the calibrated Bonner Sphere Spectrometer described in \cite{Tonino}. The calibration factor obtained is: $(1.33\pm0.08)10^{-4}$ V cm$^{2}$s.
\newline
Figure \ref{EPI_signal} shows the result of a measurement taken with the EPI3 detector placed inside the e\_LiBANS epithermal cavity (figure \ref{fig:i}) with the Linac operated at a rate of 400 Monitor Unit (MU)/min.
Diode2 and Diode1 signals are shown independently. Diode1 signal amplitudes are mainly concentrated in the lower part of the spectrum below 1.5 V while Diode2 signal amplitudes extend to higher values up to 3 V. By properly setting a threshold a Region Of Interest (ROI) can be defined, in order to maximize the neutron signal and to minimize the background contribution.
\newline
To properly define the ROI, the following R factor has been computed as a function of the threshold voltage
\begin{equation}
    R = \frac{Diode 2 - Diode 1}{Diode 2}
\end{equation}
\begin{figure}[htbp]
\centering
    \includegraphics[width=.47\textwidth,origin=c]{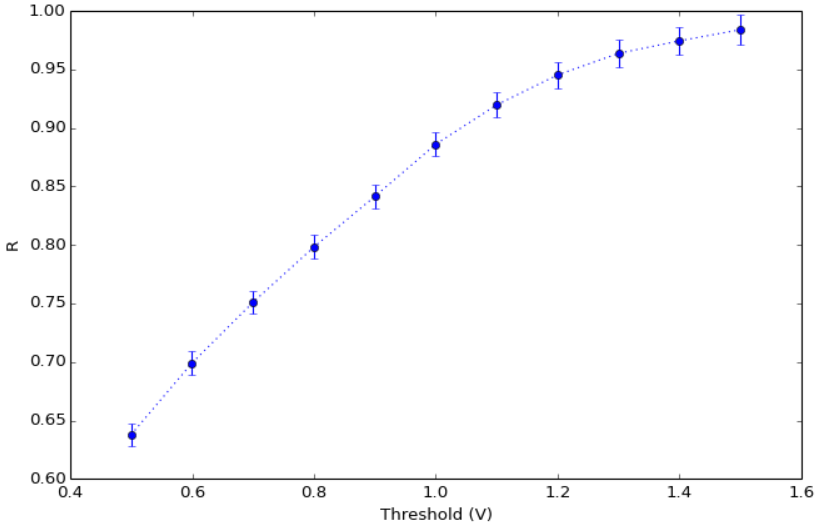}
    \qquad
    \includegraphics[width=.41\textwidth,origin=c]{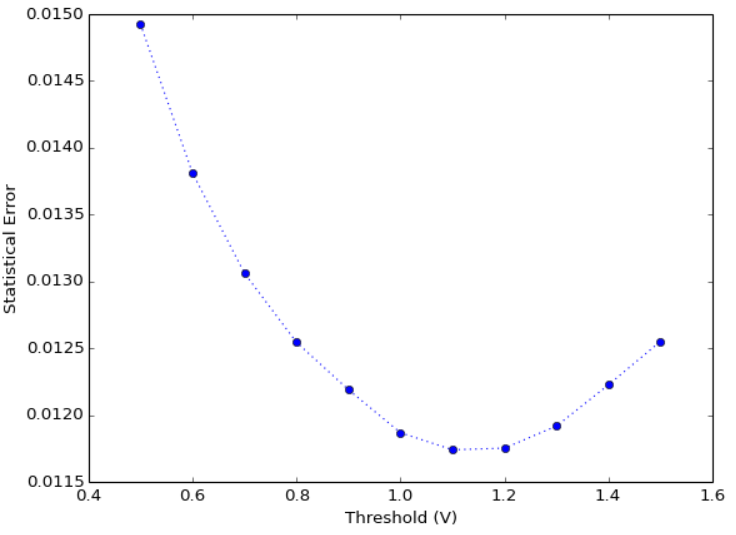}
    \caption{{\bf Left}: Dependence of the R ratio on the threshold voltage. {\bf Right}: Statistical error on the difference between the two diode signals as a function of the threshold voltage.}
    \label{fig:EPI_rates}
\end{figure}
\newline
The dependence of the R ratio on the threshold voltage is shown in figure \ref{fig:EPI_rates}.
A threshold value of 1.1 V has been chosen corresponding to R = $0.91\pm0.01$, so that the statistical error caused by the subtraction of the two diode signals is minimized .
\newline
The EPI3 detector has then been used to measure the epithermal neutron fluence rate in the e\_LiBANS cavity operating the LINAC at 439 MU/min. The epithermal neutron fluence rate extracted from EPI3 measurement is $(2.98\pm0.23)10^{4}$ cm$^{-2}$s$^{-1}$.
\newline
In order to check the EPI3 linearity the measurement the has been repeated at different LINAC rates. The result is shown in figure \ref{fig:EPI3_linearita}.

\newpage
\section{Conclusions}

The thermal and epithermal accelerator based neutron sources developed at the Physics Department of the University of Torino, together with the HOTNES and EPINES sources in Frascati, offer small scale, easily accessible irradiation facilities in which it is  possible to test and to calibrate novel diagnostics.  
\newline
This paper shows the results that have been obtained for the following novel active devices:
\begin{enumerate}
    \item \textit{Silicon Carbide neutron detector}\newline
    Starting from a commercial silicon carbide substrate, a thermal neutron fluence meter has been produced. Its sensitive area can be properly scaled with neutron field intensity. This novel device has proved to be gamma insensitive and able to operate in pulsed fields, with a linear response to the thermal neutron fluence rate. Concerning its radiation hardness it has proved to be still operable and linear after being exposed to an integrated neutron fluence of $10^{13}$ cm$^{-2}$, although with reduced efficiency. Based on 4 x 4 SiC devices a beam monitor detector has been assembled covering a field of view of 10 x 10 cm$^2$ and its use has been demonstrated by mapping the transverse field inside the e\_LiBANS cavity.  
    \item \textit{Epithermal Neutron Rate Detector}\newline
    Following the original design of the TNRD, though optimised for the epithermal range, a compact, linear, gamma insensitive, epithermal neutron fluence meter, able to operate in a pulsed neutron field, has been produced. 
    Its reduced dimensions make it suitable to be used for carefully mapping epithermal neutron fields. 
    The epithermal neutron rate at the e\_LiBANS facility measured with this device turned out to be  $(2.98\pm0.23)10^{4}$ cm$^{-2}$s$^{-1}$, in agreement with the expected value from the MCNP6 simulations.
\end{enumerate} 

\acknowledgments

This project has been supported by Compagnia di San Paolo grant "OPEN ACCESS LABS" (2015), Fondazione CRT grant n.2015.AI1430.U1925, INFN CSN 5, MIUR Dipartimenti di Eccellenza (ex L. 232/2016, art. 1, cc. 314, 337). The authors are also grateful to Azienda Sanitaria Ospedaliera San Luigi Gonzaga - Orbassano, Ospedale S. Giovani Antica Sede - Torino and to the Elekta S.p.A. for the technical support in the LINAC comissioning and maintenance.


\end{document}